\documentstyle[12pt,graphicx,pstricks,braket,amsmath,amssymb,enumitem,
                   float, epstopdf,pdfpages,cite,tabu,hyperref,subfigure, float,tikz, yfonts]{article}
                    \usetikzlibrary{arrows,matrix,positioning}
\numberwithin{equation}{section}

\newcommand{\hi}[1]{}

\DeclareMathOperator{\Tri}{tri}
\begin{document}

\def\AEF{A.E. Faraggi}

\def\JHEP#1#2#3{{JHEP} {\textbf #1}, (#2) #3}
\def\vol#1#2#3{{\bf {#1}} ({#2}) {#3}}
\def\NPB#1#2#3{{\it Nucl.\ Phys.}\/ {\bf B#1} (#2) #3}
\def\PLB#1#2#3{{\it Phys.\ Lett.}\/ {\bf B#1} (#2) #3}
\def\PRD#1#2#3{{\it Phys.\ Rev.}\/ {\bf D#1} (#2) #3}
\def\PRL#1#2#3{{\it Phys.\ Rev.\ Lett.}\/ {\bf #1} (#2) #3}
\def\PRT#1#2#3{{\it Phys.\ Rep.}\/ {\bf#1} (#2) #3}
\def\MODA#1#2#3{{\it Mod.\ Phys.\ Lett.}\/ {\bf A#1} (#2) #3}
\def\RMP#1#2#3{{\it Rev.\ Mod.\ Phys.}\/ {\bf #1} (#2) #3}
\def\IJMP#1#2#3{{\it Int.\ J.\ Mod.\ Phys.}\/ {\bf A#1} (#2) #3}
\def\nuvc#1#2#3{{\it Nuovo Cimento}\/ {\bf #1A} (#2) #3}
\def\RPP#1#2#3{{\it Rept.\ Prog.\ Phys.}\/ {\bf #1} (#2) #3}
\def\APJ#1#2#3{{\it Astrophys.\ J.}\/ {\bf #1} (#2) #3}
\def\APP#1#2#3{{\it Astropart.\ Phys.}\/ {\bf #1} (#2) #3}
\def\EJP#1#2#3{{\it Eur.\ Phys.\ Jour.}\/ {\bf C#1} (#2) #3}
\def\etal{{\it et al\/}}
\def\notE6{{$SO(10)\times U(1)_{\zeta}\not\subset E_6$}}
\def\E6{{$SO(10)\times U(1)_{\zeta}\subset E_6$}}
\def\highgg{{$SU(3)_C\times SU(2)_L \times SU(2)_R \times U(1)_C \times U(1)_{\zeta}$}}
\def\highSO10{{$SU(3)_C\times SU(2)_L \times SU(2)_R \times U(1)_C$}}
\def\lowgg{{$SU(3)_C\times SU(2)_L \times U(1)_Y \times U(1)_{Z^\prime}$}}
\def\SMgg{{$SU(3)_C\times SU(2)_L \times U(1)_Y$}}
\def\Uzprime{{$U(1)_{Z^\prime}$}}
\def\Uzeta{{$U(1)_{\zeta}$}}

\newcommand{\cc}[2]{c{#1\atopwithdelims[]#2}}
\newcommand{\bev}{\begin{verbatim}}
\newcommand{\beq}{\begin{equation}}
\newcommand{\ba}{\begin{eqnarray}}
\newcommand{\ea}{\end{eqnarray}}

\newcommand{\beqa}{\begin{eqnarray}}
\newcommand{\beqn}{\begin{eqnarray}}
\newcommand{\eeqn}{\end{eqnarray}}
\newcommand{\eeqa}{\end{eqnarray}}
\newcommand{\eeq}{\end{equation}}
\newcommand{\beqt}{\begin{equation*}}
\newcommand{\eeqt}{\end{equation*}}
\newcommand{\Eev}{\end{verbatim}}
\newcommand{\bec}{\begin{center}}
\newcommand{\eec}{\end{center}}
\newcommand{\bes}{\begin{split}}
\newcommand{\ees}{\end{split}}
\def\ie{{\it i.e.~}}
\def\eg{{\it e.g.~}}
\def\half{{\textstyle{1\over 2}}}
\def\nicefrac#1#2{\hbox{${#1\over #2}$}}
\def\third{{\textstyle {1\over3}}}
\def\quarter{{\textstyle {1\over4}}}
\def\m{{\tt -}}
\def\mass{M_{l^+ l^-}}
\def\p{{\tt +}}

\def\slash#1{#1\hskip-6pt/\hskip6pt}
\def\slk{\slash{k}}
\def\GeV{\,{\rm GeV}}
\def\TeV{\,{\rm TeV}}
\def\y{\,{\rm y}}

\def\l{\langle}
\def\r{\rangle}
\def\LRS{LRS  }

\begin{titlepage}
\samepage{
\setcounter{page}{1}
\rightline{}
\vspace{1.5cm}

\begin{center}
 {\Large \bf Free-Fermionic $SO(8)$ And $\Tri({\mathbb{O}})$}
\end{center}

\begin{center}

{\large
Johar M. Ashfaque$^\spadesuit$\footnote{email address: jauhar@liv.ac.uk}
}\\
\vspace{1cm}
$^\spadesuit${\it  Dept.\ of Mathematical Sciences,
             University of Liverpool,
         Liverpool L69 7ZL, UK\\}
\end{center}

\begin{abstract}
In this note, we speculate about the fundamental role being played by the $SO(8)$ group representations displaying the triality structure that necessarily arise in models constructed under the free fermionic methodology as being remnants of the higher-dimensional triality algebra 
$$\Tri({\mathbb{O}}) = {\mathfrak{so }} (8).$$
\end{abstract}
\smallskip}
\end{titlepage}

\section{Introduction}
 
The heterotic $E_8 \times E_8$ and the heterotic $SO(32)$ hold a special place when it comes to relating string vacua to experimental phenomena. In this note, we speculate about the fundamental role being played by the $SO(8)$ group representations, displaying the triality structure which is the four dimensional manifestation of the twisted generation of gauge groups already
noticed in the ten dimensional case, that necessarily arise in models constructed following the free fermionic methodology being remnants of the higher-dimensional triality algebra 
$$\Tri({\mathbb{O}}) = {\mathfrak{so }} (8).$$
 
\section{{The Free-Fermionic Methodology}}
For each consistent heterotic string model, there exists a partition function defined by a set of vectors with boundary conditions and a set of coefficients associated to each pair of these vectors. It will be shown that for each set of boundary conditions basis vectors and the set of associated coefficients, a set of general rules can be summarized for any model realized in the free fermionic formalism. These rules, originally derived by Antoniadis, Bachas, Kounnas in \cite{fff}, are known as the ABK rules\footnote{These rules were also developed with a different formalism by Kawai, Lewellen and Tye in \cite{fff1}.}
For further convenience, the vectors containing the boundary conditions used to define a model are called the basis vectors and the associated coefficients are called the one-loop phases that appear in the partition function. 
\subsection{{The ABK Rules}}
One of the key elements is the set of basis vectors that defines $\Xi$, the space of all the sectors. For each sector $\beta \in \Xi$ there is a corresponding Hilbert space of states. Each basis vector $b_{i}$ consists of a set of boundary conditions for each fermion denoted by
$$b_{i}=\{\alpha(\psi^{\mu}_{1,2}), ...,\alpha(\omega^{6})|\alpha(\overline y^{1}),...,\alpha(\overline{\phi}^{8})\}$$
where $\alpha(f)$ is defined by
$$f \rightarrow - e^{\ i \pi \alpha(f)}f.$$
The $b_{i}$ have to form an additive Abelian group and satisfy the constraints. If $N_i$ is the smallest positive integer for which $N_{i}b_i =0$ and $N_{ij}$ is the least common multiple of $N_i$ and $N_j$ then the rules for the basis vectors, known popularly as the ABK rules, are given as
\begin{align}
	&(1) \quad \sum m_i \cdot b_i = 0 \iff m_{i}=0 \mod N_{i}\,\,\forall i
	\\
	&(2) \quad N_{ij} \cdot  b_i \cdot b_j = 0 \mbox{ mod } 4
	\\
	&(3) \quad N_{i} \cdot  b_i \cdot b_i = 0 \mbox{ mod } 8
	\\
	&(5) \quad b_1 = \mathbf{1}\iff \mathbf{1}\in \Xi
	\\
	&(4) \quad \mbox{Even number of real fermions}
\end{align}
where
$$
b_i \cdot b_j = \left ( \frac{1}{2} \sum_{\mbox{left real}} + \sum_{\mbox{left complex}} - \frac{1}{2} \sum_{\mbox{right real}} - \sum_{\mbox{right complex}} \right ) b_i (f) \times b_j(f).
$$
\subsection{{Rules for the One-Loop Phases}}
The rules for the one-loop phases are
\begin{eqnarray}
	C 
	\binom{b_{i}}{b_{j}}
	&=& \delta_{b_j} e^{\frac{2i \pi}{N_j}n} = \delta_{b_i} e^{\frac{2i \pi}{N_i}m} e^{i \pi \frac{b_i \cdot b_j}{N_j}n}
	\\
	C 
	\binom{b_{i}}{b_{i} }
	&=& -e^{\frac{i \pi}{4} b_i \cdot b_j}
	C
	\binom{b_{i}}{1}\\
	C 
	\binom{b_{i}}{b_{j}}
	&=& e^{\frac{i \pi}{2} b_i \cdot b_j}
	C
	\binom{b_{i}}{1}^*
	\\
	C 
	\binom{b_{i}}{b_{j} + b_{k}}
	&=& \delta_{b_i} 
	C
	\binom{b_{i}}{b_{j}}
	C
	\binom{b_{i}}{b_{k}}
\end{eqnarray}
where the spin-statistics index is defined as
\[
\delta_{\alpha} = e^{i \alpha(\psi^\mu) \pi} = \begin{cases} 
\begin{array}{cc}
\,\,\,\,1, \quad \quad &\alpha(\psi_{1,2}) = 0 \\
-1, \quad \quad &\alpha(\psi_{1,2}) = 1 \\
\end{array}\end{cases}.
\]

\subsection{{The GGSO Projections}}
To complete this construction, we have to impose another set of constraints on the physical states called the GGSO projections. The GGSO projection selects the states ${|S\rangle}_\alpha$ belonging to the $\alpha$ sector satisfying
\begin{equation}
	e^{i \pi b_i \cdot F_\alpha} {|S\rangle}_\alpha = \delta_\alpha C
	\binom{\alpha}{b_{i}}^* 
	{|S\rangle}_\alpha
	\quad \quad
	\
	\quad \quad
	\forall\,\, b_i
\end{equation}
where 
\begin{equation}
	b_i \cdot F_\alpha = \left ( \frac{1}{2} \sum_{\mbox{left real}} + \sum_{\mbox{left complex}} - \frac{1}{2} \sum_{\mbox{right real}} - \sum_{\mbox{right complex}} \right ) b_i (f)\times F_\alpha(f)
\end{equation}
where $F_\alpha(f)$ is the fermion number operator given by
\[
F_\alpha(f) =  \begin{cases}
\begin{array}{cc}
+1, \quad \quad & \mbox{if } f \mbox{ is a fermionic oscillator,}\\
-1, \quad \quad &\mbox{if } f \mbox{ is the complex conjugate.} 
\end{array}\end{cases}
\]
\subsection{{The Massless String Spectrum}}
As we are interested in low-energy physics, we are only interested in the massless states. The physical states in the string spectrum satisfy the
level matching condition
\beq
M_L^2=-{1\over 2}+{{{\alpha_L}\cdot{\alpha_L}}\over 8}+N_L=-1+
{{{\alpha_R}\cdot{\alpha_R}}\over 8}+N_R=M_R^2
\label{virasorocond}
\eeq
where
$\alpha=(\alpha_L;\alpha_R)\in\Xi$ is a sector in the additive group, and
\beq
N_L=\sum_f ({\nu_L}) ;\hskip 3cm N_R=\sum_f ({\nu_R});
\label{nlnr}
\eeq
The frequencies of the fermionic oscillators depending on their boundary conditions is taken to be 
$$f \rightarrow - e^{\ i \pi \alpha(f)}f, \qquad f^{\ast}\rightarrow - e^{-i \pi \alpha(f)}f^{\ast}.$$
The frequency for the fermions is given by
$$\nu_{f, f^{\ast}} = \frac{1\pm \alpha(f)}{2}.$$
Each complex fermion $f$ generates a $U(1)$ current with a charge with respect to the unbroken Cartan generators of the four dimensional gauge group given by
\begin{eqnarray*}
	Q_{\nu}(f)&=&\nu -\frac{1}{2}\\
	&=& \frac{\alpha(f)}{2}+F
\end{eqnarray*}
for each complex right--moving fermion $f$.
\subsection{{The Enhancements}}
Extra space-time vector bosons may be obtained from the sectors satisfying the conditions:
$$\alpha_{L}^{2}=0,\qquad\qquad \alpha_{R}^{2}\neq0.$$
There are three possible types of enhancements:
\begin{itemize}
	\item Observable for example $x$,
	\item Hidden for example $z_{1}+z_{2}$,
	\item Mixed for example $\alpha$.
\end{itemize}

\section{{The Free-Fermionic $4D$ Models}}
The phenomenological free fermionic heterotic string models were
constructed following two main routes, the first are the so called
NAHE--based models.
This set of models utilise a set of eight or nine
boundary condition basis vectors. The first five consist of the so--called
NAHE set \cite{nahe} and are common in all these models. The basis vectors
underlying the NAHE--based models therefore differ by the additional
three or four basis vectors that extend the NAHE set.

\noindent The second route follows from the classification methodology that was
developed in \cite{gkr} for the classification of type II
free fermionic superstrings and adopted in
\cite{fknr, fkr, acfkr, frs} for the
classification of free fermionic
heterotic string vacua with $SO(10)$ GUT symmetry and its Pati--Salam
\cite{acfkr} and Flipped $SU(5)$ \cite{frs} subgroups. The main
difference between the two classes of models is that while the
NAHE--based models allow for asymmetric boundary conditions with respect
to the set of internal fermions $\{ y, \omega\vert {\bar y}, {\bar\omega}\}$,
the classification method only utilises symmetric
boundary conditions. This distinction affects the moduli spaces
of the models \cite{moduli}, which can be entirely fixed in the
former case \cite{cleaver} but not in the later.
On the other hand the classification method enables the systematic
scan of spaces of the order of $10^{12}$ vacua, and led to the
discovery of spinor--vector duality \cite{fkr, svduality} and
exophobic heterotic string vacua \cite{acfkr}. 

\subsection{The Classification Methodology}
A subset of basis vectors
that respect the $SO(10)$ symmetry is given by
the set of 12 boundary condition basis vectors
$V=\{v_1,v_2,\dots,v_{12}\}
$
\begin{eqnarray*}
	v_1=&1&=\{\psi_{\mu}^{1,2}, \chi^{1,...,6}, y^{1,...,6}, \omega^{1,...,6}|\bar{y}^{1,...,6}, \bar{\omega}^{1,...,6}, \bar{\psi}^{1,...,5}, \bar{\eta}^{1,2,3}, \bar{\phi}^{1,...,8}\}, \\
	v_2=&S&=\{\psi^\mu,\chi^{12},\chi^{34},\chi^{56}\},\nonumber\\
	v_{2+i}=&e_i&=\{y^{i},\omega^{i}|\overline{y}^i,\overline{\omega}^i\}, \
	i=1,\dots,6,\nonumber\\
	v_{9}=&b_{1}&=\{\chi^{34},\chi^{56},y^{34},y^{56}|\bar{y}^{34},
	\overline{y}^{56},\overline{\eta}^1,\overline{\psi}^{1,\dots,5}\},\label{basis}\\
	v_{10}=&b_{2}&=\{\chi^{12},\chi^{56},y^{12},y^{56}|\overline{y}^{12},
	\overline{y}^{56},\overline{\eta}^2,\overline{\psi}^{1,\dots,5}\},\nonumber\\
	v_{11}=&z_1 &=\{ \overline{\phi}^{1,...,4}\},\nonumber\\
	v_{12}=&z_2&= \{ \overline{\phi}^{5,...,8}\}\nonumber\\
\end{eqnarray*}
where the basis vectors ${{1}}$ and ${{S}}$, generate a model
with the $SO(44)$ gauge symmetry and ${N} = 4$
space--time SUSY with the tachyons being projected out of the massless spectrum. The next six basis vectors: $e_{1},...,e_{6}$ all correspond to the possible symmetric shifts of the six internal coordinates thus breaking the $SO(44)$ gauge group to $SO(32)\times U(1)^{6}$ but keeping the ${N}=4$ SUSY intact. The vectors $b_{i}$ for ${{i}}=1,2$ correspond to the ${\mathbb{Z}}_2 \times {\mathbb{Z}}_2$ orbifold twists. The vectors ${b_{1}}$ and ${b_{2}}$ play the role of breaking the ${N}=4$  down to ${N}=1$ whilst reducing the gauge group to $SO(10)\times U(1)^{2}\times SO(18)$. The states coming from the hidden sector are produced by ${z_{1}}$ and  ${z_{2}}$ left untouched by the action of previous basis vectors. These vectors together with the others generate the following adjoint representation of the gauge symmetry: $SO(10)\times U(1)^{3}\times SO(8)\times SO(8)$ where $SO(10)\times U(1)^{3}$ is the observable gauge group which gives rise to matter states from the twisted sectors charged under the $U(1)$s while $SO(8)\times SO(8)$ is the hidden gauge group gives rise to matter states which are neutral under the $U(1)$s.
\subsection{{The Various $SO(10)$ Subgroups}}
The $SO(10)$ GUT models generated can be broken to one of its subgroups by the boundary condition assignment on the complex fermion $\overline{\psi}^{1,...,5}$. For the Pati-Salam and the Flipped $SU(5)$ case, one additional basis vector is required to break the $SO(10)$ GUT symmetry. However, in order to construct the $SU(4)\times SU(2)\times U(1)$, the Standard-Like models and the Left-Right Symmetric models, the Pati-Salam breaking is required along with an additional $SO(10)$ breaking basis vector. The following boundary condition basis vectors can be used to construct the necessary gauge groups: 
\subsubsection{{The Pati-Salam Subgroup}}
$$v_{13}=\alpha=\{\overline{\psi}^{4,5}, \overline{\phi}^{1,2}\}$$
\subsubsection{{The Flipped $SU(5)$ Subgroup}}
$$v_{13}=\alpha=\{\overline{\eta}^{1,2,3}=\frac{1}{2}, \overline{\psi}^{1,...,5}=\frac{1}{2}, \overline{\phi}^{1,...,4}=\frac{1}{2}, \overline{\phi}^{5}\}$$
\subsubsection{{The $SU(4)\times SU(2)\times U(1)$ Subgroup}}
\begin{eqnarray*}
	v_{13}&=\alpha&= \{\overline{\psi}^{4,5}, \overline{\phi}^{1,2}\}\\
	v_{14}&=\beta&= \{\overline{\psi}^{4,5}=\frac{1}{2}, \overline{\phi}^{1,...,6}=\frac{1}{2}\}
\end{eqnarray*}
\subsubsection{{The Left-Right Symmetric Subgroup}}
\begin{eqnarray*}
	v_{13}=\alpha &=& \{ \overline{\psi}^{4,5}, \overline{\phi}^{1,2}\},\nonumber\\
	v_{14}=\beta &=& \{ \overline{\eta}^{1,2,3}=\textstyle\frac{1}{2},\overline{\psi}^{1,...,3}=\frac{1}{2},\overline{\phi}^{1,2}=\frac{1}{2},\overline{\phi}^{3,4}\}\nonumber\\
\end{eqnarray*}

\subsubsection{{The Standard-Like Model Subgroup}}
\begin{eqnarray*}
	v_{13}&=\alpha&= \{\overline{\psi}^{4,5}, \overline{\phi}^{1,2}\}\\
	v_{14}&=\beta&= \{\overline{\eta}^{1,2,3}=\frac{1}{2}, \overline{\psi}^{1,...,5}=\frac{1}{2}, \overline{\phi}^{1,...,4}=\frac{1}{2}, \overline{\phi}^{5}\}
\end{eqnarray*}

\section{The $SU421$ And LRS Models}
In \cite{Faraggi:2015iaa}, the fact was highlighted that the LRS and $SU421$ models are not viable as these models circumvent the $E_6 \rightarrow SO(10)\times U(1)_{\zeta}$ symmetry breaking pattern with the price that the $U(1)_{\zeta}$ charges of the SM states do not satisfy the $E_6$ embedding necessary for unified gauge couplings to agree with the low energy values of $\sin^2 \theta_W (M_{Z})$ and $\alpha_{s} (M_Z)$ \cite{Faraggi:2011xu, Faraggi:2013nia}. 

While the statement is true for the $SU421$ models \cite{Cleaver:2002ps, Faraggi:2014vma}, we introduce LRS models which are constructed in the free fermionic formalism to verify that the $U(1)_{\zeta}$ charges of the SM states satisfy the $E_6$ embedding. We begin by assuming that $U(1)_\zeta$ 
charges admit the $E_6$ embedding. In this case the heavy Higgs states consists of the pair ${\cal N}\left({\bf1},{\bf\frac{3}{2}},{\bf1},{\bf2},{\bf\frac{1}{2}}\right),~
{\bar{\cal N}}\left({\bf1},-{\bf\frac{3}{2}},{\bf1},{\bf2}, 
-{\bf\frac{1}{2}}\right).
$
The VEV along the electrically neutral component leaves unbroken
the SM gauge group and the $U(1)_{Z^\prime}$ combination  \beq
U(1)_{{Z}^\prime} ~=~
{1\over {2}} U(1)_{B-L} -{2\over3} U(1)_{T_{3_R}} - {5\over3}U(1)_\zeta
~\notin~ SO(10)\nonumber
\eeq
where $U(1)_{\zeta}=\sum_{i=1}^{3}U(1)_i$ is anomaly free which may remain unbroken down to low scales. 
We remark, however, that in the NAHE-based free fermionic LRS models \cite{lrs} 
the $U(1)_\zeta$ charges do not admit the $E_6$ embedding and go on to show that the same is true for free fermionic models constructed by utilizing the classification methodology \cite{gkr}.

\subsection{{The Non-Viable $SU(4)\times SU(2)\times U(1)$}}
In this section, we briefly consider the model presented in \cite{Faraggi:2014vma} which was obtained using the classification methodology. The set of basis vectors that generate the  $SU(4)\times SU(2)\times U(1)$ heterotic string model are given by
\begin{eqnarray*}
	v_1=&1&=\{\psi_{\mu}^{1,2}, \chi^{1,...,6}, y^{1,...,6}, \omega^{1,...,6}|\bar{y}^{1,...,6}, \bar{\omega}^{1,...,6}, \bar{\psi}^{1,...,5}, \bar{\eta}^{1,2,3}, \bar{\phi}^{1,...,8}\}, \\
	v_2=&S&=\{\psi^\mu,\chi^{12},\chi^{34},\chi^{56}\},\nonumber\\
	v_{2+i}=&e_i&=\{y^{i},\omega^{i}|\overline{y}^i,\overline{\omega}^i\}, \
	i=1,\dots,6,\nonumber\\
	v_{9}=&b_{1}&=\{\chi^{34},\chi^{56},y^{34},y^{56}|\bar{y}^{34},
	\overline{y}^{56},\overline{\eta}^1,\overline{\psi}^{1,\dots,5}\},\label{basis}\\
	v_{10}=&b_{2}&=\{\chi^{12},\chi^{56},y^{12},y^{56}|\overline{y}^{12},
	\overline{y}^{56},\overline{\eta}^2,\overline{\psi}^{1,\dots,5}\},\nonumber\\
	v_{11}=&z_1 &=\{ \overline{\phi}^{1,...,4}\},\nonumber\\
	v_{12}=&z_2&= \{ \overline{\phi}^{5,...,8}\}\nonumber\\
	v_{13}=&\alpha&= \{\overline{\psi}^{4,5}, \overline{\phi}^{1,2}\},\\
	v_{14}=&\beta&= \{\overline{\psi}^{4,5}=\frac{1}{2}, \overline{\phi}^{1,...,6}=\frac{1}{2}\}
\end{eqnarray*}
where the space-time vector bosons are obtained solely from the untwisted
sector and generate the observable and hidden gauge 
symmetries, given by:
\beqn
{\rm observable} ~: &~~~~~~~~SU(4)\times SU(2)\times U(1)\times U(1)^3 \nonumber\\
{\rm hidden}     ~: &SU(2)\times U(1)\times SU(2)\times U(1)\times SU(2)\times U(1)\times SO(4)\nonumber
\eeqn
In order to preserve the aforementioned observable and hidden gauge groups, all the additional space–time vector bosons need to be
projected out which can arise from the following $36$ sectors as enhancements:
$$\begin{Bmatrix}
z_{1},&z_{1}+\beta,&z_{1}+2\beta,\\
z_{1}+\alpha,&z_{1}+\alpha+\beta,&z_{1}+\alpha+2\beta,\\
z_{2},&z_{2}+\beta,&z_{2}+2\beta,\\
z_{2}+\alpha,&z_{2}+\alpha+\beta,&z_{2}+\alpha+2\beta,\\
z_{1}+z_{2},&z_{1}+z_{2}+\beta,&z_{1}+z_{2}+2\beta,\\
z_{1}+z_{2}+\alpha,&z_{1}+z_{2}+\alpha+\beta,&z_{1}+z_{2}+\alpha+2\beta,\\
\beta,& 2\beta, & \alpha,\\
\alpha+\beta, &\alpha+2\beta, & x\\
z_{1}+x+\beta,&z_{1}+x+2\beta, &z_{1}+x+\alpha,\\
z_{1}+x+\alpha+\beta,&z_{2}+x+\beta,&z_{2}+x+\alpha+\beta,\\
z_{1}+z_{2}+x+\beta, &z_{1}+z_{2}+x+2\beta,&  z_{1}+z_{2}+x+\alpha+\beta\\
x+\beta, & x+\alpha, &x+\alpha+\beta,\\
\end{Bmatrix}$$
where $x=\{\overline{\psi}^{1,...,5},\overline{\eta}^{1,2,3}\}$. The conclusion was reached that the $SU421$ class of models is the only class that is excluded in vacua with symmetric internal boundary conditions.

\subsection{The Free Fermionic LRSz Model Gauge Group}\label{1z}
In this section, we present the LRS model constructed using the free-fermionic construction with one $z$ basis vector. This model is generated by the following set of basis vectors:
\begin{eqnarray*}
	v_1=S&=&\{\psi^\mu,\chi^{12},\chi^{34},\chi^{56}\},\nonumber\\
	v_{1+i}=e_i&=&\{y^{i},\omega^{i}|\overline{y}^i,\overline{\omega}^i\}, \
	i=1,\dots,6,\nonumber\\
	v_{8}=b_1&=&\{\chi^{34},\chi^{56},y^{34},y^{56}|\bar{y}^{34},
	\overline{y}^{56},\overline{\eta}^1,\overline{\psi}^{1,\dots,5}\},\label{basis}\\
	v_{9}=b_2&=&\{\chi^{12},\chi^{56},y^{12},y^{56}|\overline{y}^{12},
	\overline{y}^{56},\overline{\eta}^2,\overline{\psi}^{1,\dots,5}\},\nonumber\\
	v_{10}=z &=& \{ \overline{\phi}^{1,...,8}\},\nonumber\\
	v_{11}=\alpha &=& \{ \overline{\psi}^{4,5}, \overline{\phi}^{1,2}\},\nonumber\\
	v_{12}=\beta &=& \{ \overline{\eta}^{1,2,3}=\textstyle\frac{1}{2},\overline{\psi}^{1,...,3}=\frac{1}{2},\overline{\phi}^{1,2}=\frac{1}{2},\overline{\phi}^{3,4}\}\nonumber\\
\end{eqnarray*}
where 
\begin{eqnarray*}
	{{1}} &=& S + \sum_{i=1}^{6}e_i +\alpha + 2\beta + z,\\
	x&=&\alpha+2\beta,\\
	b_3 &=& b_1 +b_2 +x.\\
\end{eqnarray*}
The space-time vector bosons are obtained solely from the untwisted
sector and generate the following observable gauge 
symmetries, given by:
\beqn
{\rm observable} ~: &~~~~~~~~SU(3)\times SU(2)_L\times SU(2)_R\times U(1)\times U(1)^3 \nonumber\\
{\rm hidden} ~: &~~~~~~~~SU(2)\times U(1)\times SO(4)\times SO(8)\nonumber\eeqn

\subsection{{The Free Fermionic LRS2z Model Gauge Group}}\label{2z}
In this section, we present the LRS model constructed using the free-fermionic construction where $z_i$ basis vectors are utilized for $i=1,2$. This model is generated by the following set of basis vectors:
\begin{eqnarray*}
	v_1=&1&=\{\psi_{\mu}^{1,2}, \chi^{1,...,6}, y^{1,...,6}, \omega^{1,...,6}|\bar{y}^{1,...,6}, \bar{\omega}^{1,...,6}, \bar{\psi}^{1,...,5}, \bar{\eta}^{1,2,3}, \bar{\phi}^{1,...,8}\}, \\
	v_2=&S&=\{\psi^\mu,\chi^{12},\chi^{34},\chi^{56}\},\nonumber\\
	v_{2+i}=&e_i&=\{y^{i},\omega^{i}|\overline{y}^i,\overline{\omega}^i\}, \
	i=1,\dots,6,\nonumber\\
	v_{9}=&b_{1}&=\{\chi^{34},\chi^{56},y^{34},y^{56}|\bar{y}^{34},
	\overline{y}^{56},\overline{\eta}^1,\overline{\psi}^{1,\dots,5}\},\label{basis}\\
	v_{10}=&b_{2}&=\{\chi^{12},\chi^{56},y^{12},y^{56}|\overline{y}^{12},
	\overline{y}^{56},\overline{\eta}^2,\overline{\psi}^{1,\dots,5}\},\nonumber\\
	v_{11}=&z_1 &=\{ \overline{\phi}^{1,...,4}\},\nonumber\\
	v_{12}=&z_2&= \{ \overline{\phi}^{5,...,8}\}\nonumber,\\
	v_{13}=&\alpha &= \{ \overline{\psi}^{4,5}, \overline{\phi}^{1,2}\},\nonumber\\
	v_{14}=&\beta &= \{ \overline{\eta}^{1,2,3}=\textstyle\frac{1}{2},\overline{\psi}^{1,...,3}=\frac{1}{2},\overline{\phi}^{1,2}=\frac{1}{2},\overline{\phi}^{3,4}\}\nonumber\\
\end{eqnarray*}
The space-time vector bosons are obtained solely from the untwisted
sector and generate the following observable gauge 
symmetries, given by:
\beqn
{\rm observable} ~: &~~~~~~~~SU(3)\times SU(2)_L\times SU(2)_R\times U(1)\times U(1)^3 \nonumber\\
{\rm hidden} ~: &~~~~~~~~SU(2)\times U(1)^{3}\times SO(8)\nonumber\eeqn

\section{Descending To {\bf{D=2}}}
In this section, compactifying the heterotic–string to two dimensions, we find that the two dimensional free fermions in the light-cone gauge are the real left-moving fermions 
$$\chi^{i}, y^{i}, \omega^{i},\qquad i=1,...,8,$$
the real right-moving fermions
$$\overline{y}^{i},\overline{\omega}^{i},\qquad i=1,...,8  $$
and the complex right-moving fermions
$$\overline{\psi}^{A},\quad A=1,...,4,\quad \overline{\eta}^{B},\quad B=0,...,3,\quad \overline{\phi}^{\alpha},\quad \alpha=1,...,8.$$
The class of models we consider will be generated by a maximal set of $7$ basis vectors defined as
\begin{eqnarray*}
	v_{1}=& 1 &= \{\chi^{i}, y^{i}, \omega^{i}|\overline{y}^{i},\overline{\omega}^{i},\overline{\psi}^{A},\overline{\eta}^{B},\overline{\phi}^{\alpha}\}, \\
	v_{2}=& H_{L} &=\{\chi^{i}, y^{i}, \omega^{i}\}, \\
 	v_{3}=& z_1 &=\{\overline{\phi}^{1,...,4}\},\\
	v_{4}=& z_2& =\{\overline{\phi}^{5,...,8}\}, \\
	v_{5}=& z_3 &=\{\overline{\psi}^{A}\}, \\
	v_{6}=& z_4 &=\{\overline{\eta}^{B}\}, \\
	v_{7}=&   z_5& =\{ \overline{y}^{1,...,4}, \overline{\omega}^{1,...,4}\}\\
\end{eqnarray*}
where 
$$z_{6} = 1+H_L+\sum_{i=1}^{5}z_{i} = \{\overline{y}^{5,...,8}, \overline{\omega}^{5,...,8}\} = \{\overline{\rho}^{5,...,8}\} .$$
The set of GGSO phases is given by
\begin{center}
	$
	\bordermatrix{~ & 1 & H_L &z_{1}&z_{2}&z_{3}&z_{4}&z_{5} \cr
	1&-1&-1&+1&+1&+1&+1&+1\cr
H_L &-1&-1&-1&-1&-1&-1&-1\cr
z_{1}&+1&-1&+1&+1&+1&+1&+1\cr
z_{2}&+1&-1&+1&+1&+1&+1&+1\cr
z_{3}&+1&-1&+1&+1&+1&+1&+1\cr
z_{4}&+1&-1&+1&+1&+1&+1&+1\cr
z_{5}&+1&-1&+1&+1&+1&+1&+1\cr}
	$
\end{center}
or simply
$$-C\binom{1}{1} = -C\binom{1}{H_L}=C\binom{1}{z_{i}} =-C\binom{H_L}{H_L}=-C\binom{H_L}{z_{i}}= C\binom{z_i}{z_{i}}=C\binom{z_i}{z_{j}}=1 $$
yielding the untwisted symmetry $$SO(8)_1\times SO(8)_2\times SO(8)_3\times SO(8)_4\times SO(8)_5 \times SO(8)_6.$$
Here our focus was on the $SO(48)$ and the dedicated GGSO phases were chosen appropriately as the following table highlights:
	\begin{table}[H]
\begin{center}

	\begin{tabular}{|c|c|c|}
		\hline
		&&\\
		$C\binom{H_{L}}{z_{i}}$&$C\binom{z_{i}}{z_{i}}$&Gauge Group\\
			&&\\
			\hline
			$-$&$+$& $SO(48)$\\
			\hline
	\end{tabular}
	\caption{\label{tableb}
		\small The configuration of the symmetry groups.}
\end{center}

\end{table}

\section{Normed Division Algebras}
In this section, we briefly discuss the normed division algebras. An algebra $A$ is a vector space equipped with a bilinear multiplication rule and a unit element.  We call $A$ a division algebra if, given $x,y \in A$ with $xy = 0$, then either $x = 0$ or $y= 0$. A normed division algebra is an algebra $A$ equipped with a positive-definite norm satisfying the condition 
$$||xy|| =||x||\,\,||y||$$
which also implies $A$ is a division algebra. There is a remarkable theorem due to Hurwitz \cite{Hur}, which states that there are only four normed division algebras: the real numbers $\mathbb{R}$, the complex numbers $\mathbb{C}$, the quaternions $\mathbb{H}$ and the octonions $\mathbb{O}$.  The algebras have dimensions $n= 1,2,4$ and $8$, respectively.  They can be constructed, one-by-one, by use of the Cayley-Dickson doubling method, starting with the reals; the complex numbers are pairs of real numbers equipped with a particular multiplication rule, a quaternion is a pair of complex numbers
and an octonion is a pair of quaternions. 

There is a Lie algebra associated with the division algebras \cite{Anastasiou:2013cya} known as the triality algebra of $A$ defined as follows
$$\Tri(A) = \{(A,B,C)|A(xy)=B(x)y+xC(y)\},\qquad A,B,C \in {\mathfrak{so }}(A),\quad x,y \in A$$
where $ {\mathfrak{so }}(A)$  is the norm-preserving algebra isomorphic to $ {\mathfrak{so }}(n)$ where $n=\dim A$. 	
We are interested primarily in the case where
$$\Tri({\mathbb{O}}) = {\mathfrak{so }} (8).$$
 The division algebras subsequently can be used to describe field theory in Minkowski space using the Lie algebra isomorphism
$${\mathfrak{so}}(1,1+n)\cong {\mathfrak{sl}}(2,A)$$
particularly 
$${\mathfrak{so}}(1,9)\cong {\mathfrak{sl}}(2,\mathbb{O}).$$

\begin{figure}[ht!]
	
	\begin{center}
		\begin{tikzpicture}[scale=.8]
		\draw (5,0) node[anchor=west]  {};
		\foreach \x in {3,...,4}
		\draw[xshift=\x cm,thick] (\x cm,0) circle (.3cm);
		\draw[xshift=8 cm,thick] (30: 17 mm) circle (.3cm);
		\draw[xshift=8 cm,thick] (-30: 17 mm) circle (.3cm);
		\foreach \y in {3.15,...,3.15}
		\draw[xshift=\y cm,thick] (\y cm,0) -- +(1.4 cm,0);
		\draw[xshift=8 cm,thick] (30: 3 mm) -- (30: 14 mm);
		\draw[xshift=8 cm,thick] (-30: 3 mm) -- (-30: 14 mm);
		\end{tikzpicture}
		
	\end{center}\caption{The Dynkin Diagram of $D_{4}$.}\end{figure}
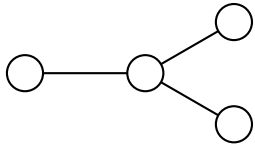

\section{Discussion}
In the free fermionic methodology the equivalence of the $8_V$, $8_{S}$
and $8_C$, $SO(8)$ representations is referred to as the triality structure. This equivalence then enables twisted constructions of the $E_8\times E_8$ or
$SO(32)$ gauge groups. The root lattice of $SO(8)$ has a quaternionic description given by the set
$$V = \bigg\{\pm1,\pm e_{1},\pm e_{2},\pm e_{3},\frac{1}{2}(\pm 1\pm e_{1}\pm e_{2}\pm e_{3})\bigg\}$$
which give the required $24$ roots. Alternatively, the root lattice of $SO(8)$ could have been composed from $SU(2)^{4}$. On the other hand, the decomposition of the adjoint representation of $E_8$ under $SO(8)\times SO(8)$ is given by 
$${\bf{248}}= ({\bf{28}},{\bf{1}})+({\bf{1}},{\bf{28}})+({\bf{8}}_v,{\bf{8}}_v)+({\bf{8}}_s,{\bf{8}}_c)+({\bf{8}}_c,{\bf{8}}_s).$$
The weights of the vectorial representation ${\bf{8}}_v$ are 
$$V_{1} = \bigg\{\frac{1}{2}(\pm1\pm e_1),\frac{1}{2}(\pm e_2\pm e_3)\bigg\},$$ 
the weights of the conjugate spinor representation ${\bf{8}}_c$ are
$$V_{2} = \bigg\{\frac{1}{2}(\pm1\pm e_2),\frac{1}{2}(\pm e_3\pm e_1)\bigg\},$$ 
and
the weights of the spinor representation ${\bf{8}}_s$ are
$$V_{3} = \bigg\{\frac{1}{2}(\pm1\pm e_3),\frac{1}{2}(\pm e_1\pm e_2)\bigg\}.$$ 
This description makes the triality of $SO(8)$ manifest. It can be easily seen that permutations of the three imaginary elements $e_1$, $e_2$ and $e_3$ will map the representations $V_1\rightarrow V_2\rightarrow V_3$. In \cite{Evans:1987tm} an explicit correspondence between simple super-Yang-Mills 
and classical superstrings in dimensions $3$, $4$, $6$, $10$ and the division 
algebras $\mathbb{R}$, $\mathbb{C}$, $\mathbb{H}$, $\mathbb{O}$ was established.

Here, we speculated about the fundamental role being played by the $SO(8)$  group representations, displaying the triality structure, which necessarily arise in models constructed under the free fermionic methodology being remnants of the higher-dimensional triality algebra, namely
$$\Tri({\mathbb{O}}) = {\mathfrak{so }} (8).$$

\section{Acknowledgements}
J. M. A. would like to thank the University of Kent and the University of Oxford for their warm hospitality.

\appendix


\begin{thebibliography}{99}
		

	\bibitem{nahe} \AEF~and D.V. Nanopoulos, \PRD{48}{1993}{3288}.
	\bibitem{fkr} A.E. Faraggi, C. Kounnas and J. Rizos,
\PLB{648}{2007}{84};
\NPB{774}{2007}{208};\\
T. Catelin-Julian, A.E. Faraggi, C. Kounnas and J. Rizos,
\NPB{812}{2009}{103}.

\bibitem{frs} \AEF, J. Rizos and H. Sonmez, \NPB{886}{2014}{202}.
\bibitem{acfkr}
B. Assel, C. Christodoulides, A.E. Faraggi, C. Kounnas and J. Rizos
\PLB{683}{2010}{306}; \NPB{844}{2011}{365}.
	\bibitem{gkr} A. Gregori, C.~Kounnas and J.~Rizos, \NPB{549}{1999}{16}.
	\bibitem{fknr} A.E. Faraggi, C. Kounnas, S.E.M. Nooij and J. Rizos,
hep-th/0311058; \NPB{695}{2004}{41}.
\bibitem{svduality} \AEF, C. Kounnas and J. Rizos, \NPB{799}{2008}{19};\\
C.~Angelantonj, A.E.~Faraggi and M.~Tsulaia,
\JHEP{1007}{2010}{004};\\
A.E.~Faraggi, I.~Florakis, T.~Mohaupt and M.~Tsulaia,
\NPB{848}{2011}{332};\\
P.~Athanasopoulos, A.E.~Faraggi and D.~Gepner,
\PLB{735}{2014}{357}.

\bibitem{moduli} \AEF, \NPB{728}{2005}{83}.

\bibitem{cleaver} G. Cleaver, \AEF, E. Manno and C. Timirgaziu,
\PRD{78}{2008}{046009}.

	\bibitem{fff}
I. Antoniadis, C. Bachas, and C. Kounnas, \NPB{289}{1987}{87};\\
I. Antoniadis and C. Bachas, \NPB{298}{1988}{586}.

\bibitem{fff1}
H. Kawai, D.C. Lewellen, and S.H.-H. Tye, \NPB{288}{1987}{1}.
\bibitem{Faraggi:2015iaa}
A.~E.~Faraggi and M.~Guzzi,
``Extra $Z^{\prime }$ s and $W^{\prime }$ s in heterotic-string derived models,''
Eur.\ Phys.\ J.\ C {\bf 75} (2015) no.11,  537
doi:10.1140/epjc/s10052-015-3763-4
[arXiv:1507.07406 [hep-ph]].
\bibitem{Faraggi:2011xu}
A.~E.~Faraggi and V.~M.~Mehta,
``Proton Stability and Light $Z^\prime$ Inspired by String Derived Models,''
Phys.\ Rev.\ D {\bf 84} (2011) 086006
doi:10.1103/PhysRevD.84.086006
[arXiv:1106.3082 [hep-ph]].
\bibitem{Faraggi:2013nia}
A.~E.~Faraggi and V.~M.~Mehta,
``Proton stability, gauge coupling unification, and a light Z′ in heterotic-string models,''
Phys.\ Rev.\ D {\bf 88} (2013) no.2,  025006
doi:10.1103/PhysRevD.88.025006
[arXiv:1304.4230 [hep-ph]].
\bibitem{lrs}
G.B. Cleaver, A.E. Faraggi and C. Savage, \PRD{63}{2001}{066001};
G.B. Cleaver, D.J Clements and A.E. Faraggi, \PRD{65}{2002}{106003};
\bibitem{Cleaver:2002ps}
G.~B.~Cleaver, A.~E.~Faraggi and S.~Nooij,
``NAHE based string models with SU(4) x SU(2) x U(1) SO(10) subgroup,''
Nucl.\ Phys.\ B {\bf 672} (2003) 64
doi:10.1016/j.nuclphysb.2003.09.012
[hep-ph/0301037].
\bibitem{Faraggi:2014vma}
A.~E.~Faraggi and H.~Sonmez,
``Classification of SU(4) X SU(2) X U(1) Heterotic-String Models,''
Phys.\ Rev.\ D {\bf 91} (2015) 066006
doi:10.1103/PhysRevD.91.066006
[arXiv:1412.2839 [hep-th]].
\bibitem{Hur}
A. Hurwitz, ``Uber die komposition der quadratishen formen von beliebig vielen variabeln,"
Nachr. Ges. Wiss. Gottingen
(1898) 309-316.
\bibitem{Anastasiou:2013cya}
A.~Anastasiou, L.~Borsten, M.~J.~Duff, L.~J.~Hughes and S.~Nagy,
``Super Yang-Mills, division algebras and triality,''
JHEP {\bf 1408} (2014) 080
doi:10.1007/JHEP08(2014)080
[arXiv:1309.0546 [hep-th]].
\bibitem{Evans:1987tm}
J.~M.~Evans,
``Supersymmetric {Yang-Mills} Theories and Division Algebras,''
Nucl.\ Phys.\ B {\bf 298} (1988) 92.
 \end{thebibliography}
\end{document}